\documentclass[twocolumn]{aastex62}
\usepackage{graphics,epsf}
\usepackage{amsmath}                
\usepackage{amsfonts}               
\usepackage{amssymb}                
\usepackage{epsfig}                 
\usepackage{graphicx}
\usepackage{float}
\usepackage{color}
\usepackage[para,online,flushleft]{threeparttable}

\newcommand{\cm}{{~\rm cm}}
\newcommand{\km}{{~\rm km}}
\newcommand{\s}{{~\rm s}}

\newcommand{\g}{{~\rm g}}

\newcommand{\K}{{~\rm K}}
\newcommand{\erg}{{~\rm erg}}
\newcommand{\yr}{{~\rm yr}}

\begin{document}

\title{Enhanced mass-loss rate evolution of stars with $\gtrsim 18 M_\odot$ and missing optically-observed type II core-collapse supernovae}

\author{Roni Anna Gofman}
\affiliation{Department of Physics, Technion, Haifa, 3200003, Israel; rongof@campus.technion.ac.il; soker@physics.technion.ac.il}

\author{Naomi Gluck}
\affiliation{Department of Physics, Stony Brook University, New York, United States; naomi.gluck@stonybrook.edu}

\author[0000-0003-0375-8987]{Noam Soker}
\affiliation{Department of Physics, Technion, Haifa, 3200003, Israel; rongof@campus.technion.ac.il; soker@physics.technion.ac.il}
\affiliation{Guangdong Technion Israel Institute of Technology, Shantou 515069, Guangdong Province, China}

\begin{abstract}
We evolve stellar models with zero-age main sequence (ZAMS) mass of $M_{\rm ZAMS} \gtrsim 18 M_\odot$ under the assumption that they experience an enhanced mass-loss rate when crossing the instability strip at high luminosities and conclude that most of them end as type Ibc supernovae (SNe Ibc) or dust-obscured SNe II. 
We explore what level of enhanced mass-loss rate during the instability strip would be necessary to explain the `red supergiant (RSG) problem'. This problem refers to the dearth of observed core-collapse supernovae progenitors with $M_{\rm ZAMS} \gtrsim 18 M_\odot$. Namely, we examine what enhanced mass loss rate could make it possible for all these stars actually to explode as CCSNe. We find that the mass-loss rate should increase by a factor of at least about ten.  We reach this conclusion by analyzing the hydrogen mass in the stellar envelope and the optical depth of the dusty wind at the explosion, and crudely estimate that under our assumptions only about a fifth of these stars explode as unobscured SNe II and SNe IIb. About 10-15 per cent end as obscured SNe II that are infrared-bright but visibly very faint, and the rest, about 65-70 per cent, end as SNe Ibc.
However, the statistical uncertainties are still too significant to decide whether many stars with $M_{\rm ZAMS} \gtrsim 18M_\odot$ do not explode as expected in the neutrino driven explosion mechanism, or whether all of them explode as CCSNe, as expected by the jittering jets explosion mechanism.
\end{abstract}


\section{Introduction}
\label{sec:intro}

The two theoretical mechanisms to power core-collapse supernova (CCSN) explosions from the gravitational energy that the collapsing core releases are: the delayed neutrino mechanism \citep{BetheWilson1985}, and the jittering jets explosion mechanism (\citealt{Soker2010}; or more generally the jet feedback mechanism, e.g., \citealt{Soker2016Rev}).
Each of these mechanisms in its originally-proposed form encounters some problems that require the addition of some ingredients.

The extra ingredient that recent numerical simulations of the delayed neutrino mechanism introduce to overcome some of the basic problems of the original delayed neutrino mechanism (for these problems see, e.g., \citealt{Papishetal2015, Kushnir2015b}), is convection above the iron core in the pre-collapse core (e.g., \citealt{CouchOtt2013, CouchOtt2015, MuellerJanka2015, Mulleretal2017, Mulleretal2019Jittering}). The flow fluctuations of the convective zone that ease explosion results in large-amplitude stochastic angular momentum variations of the mass that the newly born neutron star (NS) accretes. These fluctuations seem to lead to the launching of a bipolar outflow with varying symmetry axis directions, namely, jittering jets \citep{Soker2019JitSim}.

Indeed, the jittering jets explosion mechanism is based on such flow fluctuations in the convective regions of the pre-collapse core or envelope \citep{GilkisSoker2014, GilkisSoker2015, Quataertetal2019}. The spiral standing accretion shock instability (SASI) and other instabilities behind the stalled shock at about $100 \km$ from the newly born NS amplify these fluctuations (for the physics of the spiral SASI see, e.g., \citealt{BlondinMezzacappa2007, Iwakamietal2014, Kurodaetal2014, Fernandez2015, Kazeronietal2017}). However, results of numerical simulations that find no stochastic accretion disks around the newly born NS brought to the recognition that neutrino heating plays a role in the jittering jets explosion mechanism \citep{Soker2018KeyRoleB, Soker2019SASI}.
In a recent study, \cite{Soker2019JitSim} analyses three-dimensional hydrodynamical simulations of CCSNe and concludes that both neutrino heating and accretion of stochastic angular momentum operate together to launch jittering jets that explode CCSNe.

One of the places where the delayed neutrino mechanism and the jittering jets explosion mechanism differ from each other is the prediction of the outcome of stars with zero age main sequence (ZAMS) mass of
$M_{\rm ZAMS} \gtrsim 18 M_\odot$. According to the delayed neutrino mechanism for most of the  masses in that range the core-collapses to form a black hole in a failed supernova, i.e., there is no explosion (e.g., \citealt{Fryer1999, Horiuchietal2014, Sukhboldetal2016, Ertletal2016, SukhboldAdams2019, Ertletal2020}), but rather only a faint transient event \citep{LovegroveWoosley2013, Nadezhin1980}. According to the jittering jets explosion mechanism, on the contrary, there are no failed CCSNe, and all of these stars do explode, even if the collapsing core forms a black hole. According to the jittering jets explosion mechanism when a black hole is formed the outer core material and then the envelope gas contains enough stochastic angular momentum (e.g., \citealt{GilkisSoker2014, GilkisSoker2015, Quataertetal2019}) to launch jets and set an energetic explosion, up to $E_{\rm exp} > 10^{52} \erg$ \citep{Gilkisetal2016Super}.

These different predictions of the two explosion mechanisms relate directly to the so-called ‘red supergiant (RSG) problem’ \citep{Smarttetal2009}, referring to the finding that the observed relative number of progenitors of CCSNe II with ZAMS masses of $M_{\rm ZAMS} \gtrsim 18 M_\odot$ is much lower than their relative number on the main sequence (e.g., \citealt{Jenningsetal2014, Williamsetal2014}; for a review see, e.g., \citealt{Smartt2015R}).
\cite{Smartt2015R} argues in his thorough review of the `red supergiant problem' that it is
consistent with the claim of the delayed neutrino mechanism that most stars of
$M_{\rm ZAMS} \gtrsim 18 M_\odot$ collapse to form black holes with no visible supernovae, but possibly a faint transient event.
\cite{Adamsetal2017} suggest that the star N6946-BH1 that erupted in 2009 \citep{Gerkeetal2015} was a failed SN event of a progenitor of $\approx 25 M_\odot$. \cite{KashiSoker2017} provide an alternative interpretation to that event based on a transient event (intermediate luminosity optical transient--ILOT) that was obscured by dust in the equatorial plane that happens to be along our line of sight.

We do note that there are claims for massive progenitors of some CCSNe, e.g., a progenitor of the type IIn SN~2010jl of mass $M_{\rm ZAMS} \gtrsim 30 M_\odot$ \citep{Smithetal2011}, and
a possible SN Ic progenitor with a mass of $M_{\rm ZAMS} \gtrsim 50 M_\odot$ \citep{VanDyketal2018}.

One possible explanation to the missing massive progenitors of CCSNe II might be an obscuration by dust (e.g., \citealt{WalmswellEldridge2012}), but one should properly calculate dust extinction in CCSNe \citep{Kochaneketal2012}.

\cite{Jencsonetal2017} claim that if the two events they studied in the infrared (IR) are CCSNe, then-current optical surveys miss $\gtrsim 18\%$ of nearby CCSNe. In a more systematic study \cite{Jencsonetal2019} find nine IR bright transients, and estimate that 5 of these events are dust-obscured CCSNe (probably obscured by dusty clouds in the host galaxy). They further estimate that optical surveys miss $\approx 40 \%$ (the range of $17-64 \%$) of all CCSNe. If holds, this might cover most (or even all) stars with $M_{\rm ZAMS} \gtrsim 18 M_\odot$, implying that these stars also explode as CCSNe.

The obscured CCSNe that  \cite{Jencsonetal2019} study are most likely obscured by dusty clouds in the host galaxy, rather by a dust circumstellar matter (CSM). 
We here raise the following question. What enhanced mass loss rate during the RSG could make the CSM of some RSG with $M_{\rm ZAMS} \ga 18 M_\odot$ sufficiently dense to obscure their explosion? We also add the related question of whether our assumed enhanced mass loss rate might bring some RSG to explode as SNe Ibc rather than SNe II or IIb, even if they are not obscured by their own CSM.  

\cite{YoonCantiello2010} already studied the process by which partial ionisation of hydrogen in the envelope causes RSG stars to strongly pulsate and lose mass at a very high rate (e.g., \citealt{Hegeretal1997}). They further discussed the possibility that this enhanced mass-loss rate of stars with $M_{\rm ZAMS} \gtrsim 19-20 M_\odot$ might explain the RSG problem, by both forming an optically thick dusty CSM and by removing most,  or even all, of the hydrogen-rich envelope and forming a SN of type Ib or Ic (Ibc) progenitor. 
We continue the idea of \cite{YoonCantiello2010} but perform somewhat different evolutionary simulations. We assume that the stars have the enhanced mass-loss rate that we require to explain the RSG problem, when they cross the continuation of the instability strip on the HR diagram when they are RSGs. We strengthen the claim of \cite{YoonCantiello2010} that such an enhanced mass-loss rate might account for RSG problem, allowing all stars to explode as CCSNe. 
There are other studies that include enhanced mass loss rate of stars during the RSG phase, but they do not compare directly to the RSG problem (e.g. \citealt{Meynetetal2015}).

In Section \ref{sec:numerical}, we describe our numerical setup, and in Section \ref{sec:results} we present the calculation of evolutionary tracks under the assumption that RSG stars that cross the instability strip have very high mass-loss rates. In Section \ref{sec:RSGproblem} we study the enhanced mass loss rate that would be necessary to obscure stars with $18 M_\odot \la M_{\rm ZAMS} \la 20 M_\odot$, and determine the role of this enhanced mass-loss rate in bringing more stars of $M_{\rm ZAMS} \gtrsim 20 M_\odot$ to explode as types IIb or Ib CCSNe. We summarise our main conclusions in section \ref{sec:summary}.
    
\section{Numerical set up}
\label{sec:numerical}
\subsection{Stellar evolution}
\label{subsec:evolution}

We evolve stellar models with ZAMS mass in the range of  $M_{\rm{ZAMS}} = 15 - 30 M_{\odot}$ using Modules for Experiments in Stellar Astrophysics (\textsc{mesa}, version 10398 \citealt{Paxton2011,Paxton2013,Paxton2015,Paxton2018}). Each model has an initial metalicity of $Z=0.02$, and evolves from the pre-main sequence stage until pre-core-collapse, which we take to be the first time the iron core has an inward velocity $\geq 1000 \km$.

We employ mixing according to a mixing-length theory \citep{Henyey1965} with $\alpha_{\rm MLT}=1.5$ in convective regions defined by the Ledoux criterion. Semiconvection is used with $\alpha_{\rm sc}=1.0$ \citep{Langer1983}. Step function convective overshooting is applied with an overshooting parameter of $0.335$ \citep{Brott2011}.

We apply wind mass-loss with the \textsc{mesa} "Dutch" mass-loss scheme for massive stars which combines results from several papers and is based on 
\cite{Glebbeek2009}. For $T_{\rm eff} > 10^4 ~\rm{K}$ and surface hydrogen abundance larger than $0.4$ the "Dutch" scheme uses \cite{Vink2001} and for surface hydrogen abundance lower than $0.4$ it uses \cite{WindWR}. In cases where $T_{\rm eff} < 10^4 ~\rm{K}$ mass-loss is treated according to \cite{deJager1988}.

We break up the evolution to two parts: inside the instability strip and outside it. Figure 1 in \cite{Georgy2013} shows the Hertzsprung Russell (HR) diagram for non-rotating models with an instability strip in the range of $2 \lesssim \log \left( L / L_\odot \right) \lesssim 5 $ and $3.5 \lesssim \log \left( T_{\rm eff} ~ \left[ \rm{K} \right] \right) \lesssim 3.8 $. From that figure we approximate that the instability strip to be in the region where
\begin{equation}
    64 \lesssim \log \left( \frac{L} { L_\odot} \right) +16.4 \log \left( \frac{T_{\rm eff}} {\left[ \rm{K} \right]} \right) \lesssim 65.
\label{eq:instabilityStrip}
\end{equation}
We extend the instability strip from \cite{Georgy2013} to higher luminosities, by a linear continuation of the two boundaries of the strip on the HR diagram. Later we show that for the stars we evolve here, the center of the instability strip continuation is at about an effective temperature of $\log (T_{\rm eff}) \simeq 3.6$. This is about the same location on the HR diagram of the pulsating stars that \cite{YoonCantiello2010} study. As \cite{YoonCantiello2010} discuss, the pulsations are driven by partial ionisation of hydrogen in the envelope. 
We further note that \cite{YoonCantiello2010} assume that these stars have pulsational-driven enhanced mass loss rate. Although they did not show it from first principles, we accept here their assumption, but consider both moderate and large mass loss rate enhancement (see section \ref{subsec:Enhanced}).
   
While the star is outside the strip we set the mass-loss scaling factor to $f_{\rm ml} = 0.8$ since the models have no rotation \citep{Maeder2001}.
When the star crosses the extended part of the instability strip from right to left on the HR diagram, namely at very late evolutionary phases, we consider one of three cases for the mass-loss rate.
In the first case, we assume that the instability strip has no special role, and we keep $f_{\rm ml} = 0.8$. The other 2 cases have enhanced mass-loss inside the instability strip. Once the model enters the strip from right to left we increase the mass-loss scaling factor to $f_{\rm ml} = 2$ in one case and to $f_{\rm ml} = 10$ in another. 

As \textsc{MESA} is a numerical simulation code based on a grid (shells) and time steps, we should show that there is a convergence, i.e., no dependence on the grid resolution for the value we use  (e.g. \citealt{Farmeretal2016}). In \textsc{MESA} there is a parameter ``max\underline{\space}dq'' that sets the maximum mass in a shell, expressed as a fraction of total mass in the grid. We compared the default value of max\underline{\space}dq$=0.01$ that we use in all our cases, to several cases that we simulated with higher resolution of max\underline{\space}dq$=10^{-3}$. We found no differences in the pre-collapse characteristics we study in this paper.
    
The timestep control parameter  "delta\underline{\space}lg\underline{\space}XH\underline{\space}cntr\underline{\space}min" sets the time step as hydrogen is consumed in the core. Setting small time steps assists in calculating the passage from the ZAMS to the terminal-age main sequence. Similar parameters control the timestep for every significant evolution stage. We reduce the timesteps at these phases by setting these parameters to be $10^{-6}$ for hydrogen and helium (as in, e.g. \citealt{Farmeretal2016}), and $10^{-5}$ for heavier elements.

\subsection{Enhanced mass loss rate}
\label{subsec:Enhanced}

Our question in this study is as follows. \textit{By what factor  should we increase the mass loss rate during the RSG phase to explain the RSG problem?} We examine here two mass loss rate enhancement factors (section \ref{subsec:evolution}). We further assume, as we discussed above, that the enhanced mass loss rate occurs only when the star crosses the instability strip from right to left, i.e., when it is very bright.

We emphasise that there is no observational justifications for this enhanced mass loss rate (e.g., \citealt{Beasoretal2020}).
As we mention in section \ref{sec:intro}, based on their observations of dust-enshrouded CCSNe,  \cite{Jencsonetal2019} estimate that $\approx 17-64 \%$ of all CCSNe are dust-enshrouded. In most (or even all) cases the obscuring dust is of ISM origin. Even if in some cases CSM obscures the CCSN, it might be that in the enhanced mass loss rate of the progenitor was due to binary interaction.  

We rather raise here a theoretical question: What should the enhanced mass loss rate of single RSG stars be for single stars to explain the RSG problem?
We find below (sections \ref{sec:results}, \ref{sec:RSGproblem}) that to form a significant number of enshrouded CCSNe we need to increase the mass loss rate of single stars in the instability strip by a factor of about 10 or somewhat larger (namely, $f_{\rm ml} \simeq 10$). Much lower values will not obscure the stars when explode, and much higher values will remove too much mass before explosion. 
Namely, we have no theoretical justification for taking $f_{\rm ml} \simeq 10$, but we rather claim that \textit{if single stars form some dust-enshrouded CCSNe,} then the enhancement factor should be about 10 (here we take an enhancement factor of 12.5).

From the theoretical side, we base our prescription for enhanced mass-loss rate in the instability strip on the results of \cite{YoonCantiello2010} who argue that RSG stars lose mass at a very high rate when they are inside the instability strip on the HR diagram.
There can be two basic regimes of the mass loss rate enhancement because of pulsations in the instability strip. In the first the effect of the pulsations is linear, like the decrease in gravity and temperature as the envelope expands to maximum radius in the pulsation cycle. In this case the mass loss rate increases by a moderate factor, which we here take to be 2.5 (for $f_{\rm ml}=2$). In the other regime the effect is non linear. For example, the lower temperature together with pulsation-driven shocks in the envelope lead to substantial extra dust formation (e.g., \citealt{Goldmanetal2017}), with a large impact on the mass loss rate. Here we take the increase of the mass loss rate in the non-linear regime to be by a factor of 12.5 (for $f_{\rm ml}=10$). As we discuss below, already in the linear regime we find a non negligible influence of the enhanced mass loss rate, that becomes quite significant in the non-linear regime. 
     
The above discussion shows that from theoretical considerations the factor of $f_{\rm ml}=10$ is arbitrary, as we have no derivation of the mass loss rate increase due to the instability. However, as we explain above, we need the factor of $f_{\rm ml} \simeq 10$ to make sure that there is a significant number of dust-enshrouded CCSNe from single-star evolution.
  
We note that \cite{Meynetetal2015} conduct RSG evolution simulations where they increase the mass loss rates by a factor of 10 and 25 relative to the standard mass-loss rates during the RSG phase. We return to the study of \cite{Meynetetal2015} in section \ref{sec:results}. \cite{Georgy2012} also enhances the RSG mass loss rate, by a factor of 3 and 10, but for lower masses than what we study, i.e., $12-15 M_\odot$. Therefore, our enhanced mass loss rate factor of up to 12.5 is not an extreme in such studies.

 Finally, we note the following new results by \cite{Beasoretal2020}. \cite{Beasoretal2020} study the mass loss rate of RSGs in two stellar clusters, and conclude that the mass loss rates they find are up to a factor of 20 lower than what current evolutionary models use. If the results of \cite{Beasoretal2020} hold by future studies, it would imply that to reach the desired mass loss rate that leads to dust-enshrouded CCSNe we would have to increase the mass loss rate relative to the value that  \cite{Beasoretal2020} infer by a factor of $\approx 100$ rather than by a factor of $\simeq 10$.  Namely, during most of the evolution the mass loss rate is low, as suggested by \cite{Beasoretal2020}, but some progenitors of CCSNe suffers very high mass loss rate just before explosion due to enhanced mass loss rate, or from binary interaction. 
Specifically, we find below that for single star to explain the RSG problem, in cases of obscured CCSNe the average mass loss rate hundreds of years before explosion should be $\dot M \ga 3 \times 10^{-5} M_\odot \yr^{-1}$.
There is a clear need for further observations and theoretical study to explore the full behavior of mass loss from RSG.

\section{Results}
\label{sec:results}
    
In this section, we focus mostly on the effect of the mass-loss rate inside the instability strip on the pre-collapse state of the stellar models. We evolve over 40 stellar models up to the point of core-collapse with 16 different values of ZAMS mass for each of the three mass-loss parameters, $f_{\rm ml}$, that we set in the instability strip. 
 
\begin{figure}
    \centering
    \includegraphics[trim = 0cm 0cm 0cm 0cm ,clip=true,width=0.45\textwidth] {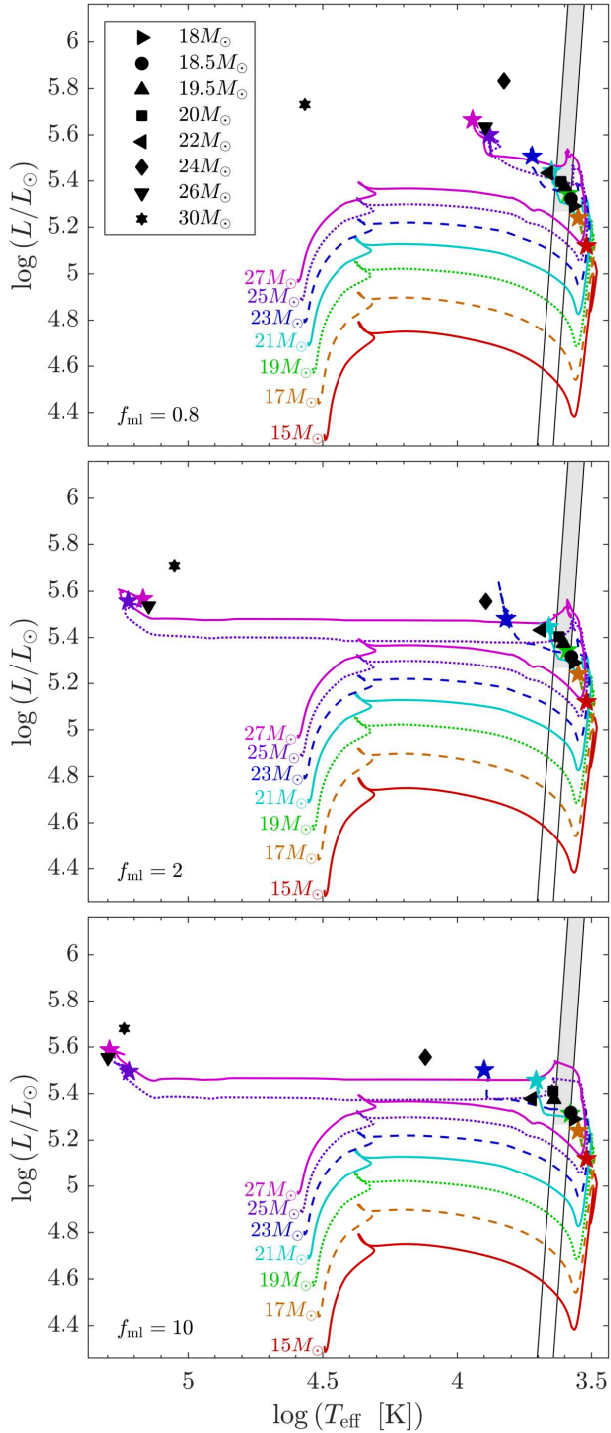} 
    \caption{The evolution track of stellar models with ZAMS masses in the range of $15-30 M_\odot$ from ZAMS to core-collapse on the HR diagram. The pre-collapse point of each model is marked by a coloured pentagram for odd masses and a black marker for all other masses. The instability strip and its extension according to equation (\ref{eq:instabilityStrip}) is marked with two black lines. The panels have different mass-loss scaling factors, $f_{\rm ml}$ as given in the inset when a star crosses the instability strip from right to left in the grey area of the strip. 
    }
    \label{fig:HRdif}
\end{figure}

Fig. \ref{fig:HRdif} shows the evolution of some models on the HR diagram, while for others we show only the final position. We also present the instability strip, including our extension to high luminosity. It is evident that by increasing the mass-loss rate when the star is inside the instability strip (extension) and crosses from right to left, the pre-collapse effective temperature of models that leave the strip increases. 

\begin{figure}
    \centering
    \includegraphics[trim= 0cm 0cm 0cm 0cm ,clip=true,width=0.45\textwidth]{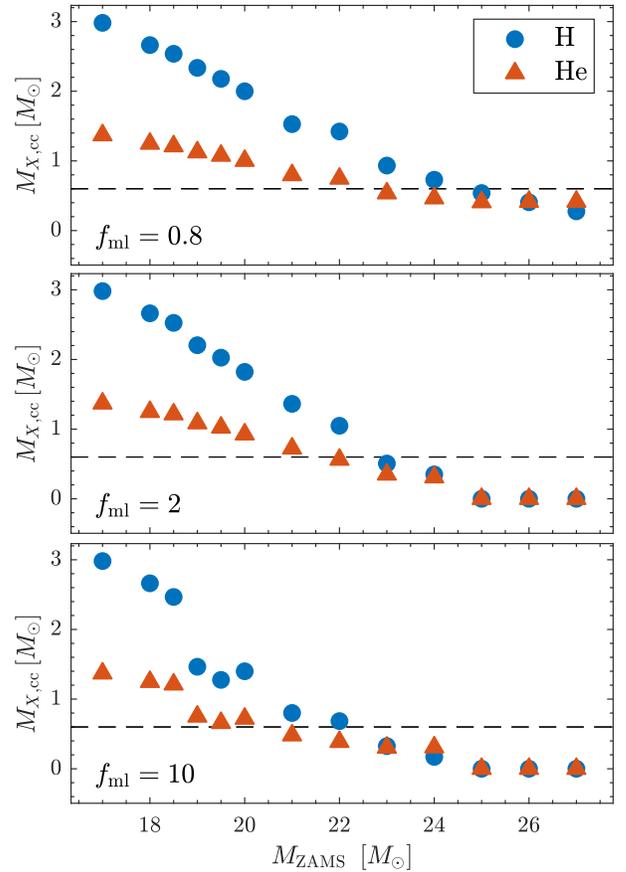}
    \caption{The final envelope mass of hydrogen (blue circles) and helium (orange triangles) as a function of the ZAMS mass. The three panels are for different mass-loss rate scaling factor, $f_{\rm ml}$, inside the instability strip as the star crosses from right to left in the HR diagram. Although the helium mass in the envelope of the most massive three models in the lower two panels is zero, there is a helium mass of about $2 M_\odot$ in the core. Therefore, these will explode as SNe Ib.
    }
    \label{fig:FinalHandHe}
\end{figure}

Moreover, models with $M_{\rm ZAMS} \gtrsim 24 M_\odot$ and enhanced mass-loss rate in the instability strip lose their entire hydrogen envelope, as we show for in Fig. \ref{fig:FinalHandHe}, and become hot progenitors of SNe Ib, i.e., Wolf-Rayet (WR) stars. Fig. \ref{fig:FinalHandHe} shows that stars with $M_{\rm ZAMS} \gtrsim 24M_\odot$ and enhanced mass-loss rate do not have hydrogen and helium in their envelope at explosion; this is because these stars loses there entire envelope by that time. These stars will explode as SNe Ib because they have a core helium layer of $\approx 2M_\odot$. Other models lose most of their hydrogen envelope but still are left with $0.01 M_\odot \la M_{\rm H, cc} \la 0.5-1 M_\odot$ of hydrogen in there envelope at core-collapse; these become the progenitors of SNe IIb. We explain the different groups and their implications on the RSG problem with more detail in section \ref{sec:RSGproblem}.
{}

Now we turn to examine the possibility of obscured CCSNe. We assume that the dense wind efficiently forms dust and calculate its optical depth. We consider the wind section from an inner radius of $R_{\rm in} = 0.5-2 \times 10^{16} \cm$, as we take the supernova to destroy dust at inner radii (calculating the exact radius requires to follow the explosion and its radiation, as well as the collision of the fastest ejecta with the dust).
We also take a density of $\rho (r) = \dot M / 4 \pi v_{\rm w} r^2$, where $\dot M$ is the mass-loss rate and $v_{\rm w}$ is the wind velocity. We also take the opacity in the V-band to be $\kappa_{\rm V} \approx 100 \cm^2 \g^{-1}$ (e.g., \citealt{Kochaneketal2012}), and derive 
\begin{equation}
\begin{aligned}
 \tau_{\rm V} =  & \intop_{R_{\rm in}} ^{R_{\rm out}} \kappa_{\rm V} \rho  ~{\rm d} r 
 \\  \simeq & ~5 
 \left( \frac {\dot M}{10^{-4} M_\odot \yr^{-1}} \right)
 \left( \frac {R_{\rm in}}{10^{16} \cm} \right)^{-1} 
 \\ & \times
 \left( \frac {\kappa_{\rm V}}{100 \cm^2 \g^{-1}} \right)
 \left( \frac {v_{\rm w}}{10 \km \s^{-1}} \right)^{-1},
\end{aligned}
\label{eq:TauV}
\end{equation}
where in the second equality we assume constant mass-loss rate and wind velocity and that $R_{\rm out} \gg R_{\rm in}$. 

To derive a more accurate expression we take the mass-loss rate as function of time from our numerical results. The density, $\rho (r)$, at radius $r$ corresponds to a mass-loss, $\dot M (t)$, at time $t=t_{\rm cc}-r/v_{\rm w}$, where $t_{\rm cc}$ is the time of core-collapse (explosion). 
We take $v_{\rm w}$ constant with time according to the following prescription. We simply assume that when the mass-loss rate in the strip is higher, the wind velocity is lower even after the star leaves the instability strip. For the default mass-loss rate, $f_{\rm ml} =0.8$, we take the wind velocity to be the escape velocity from the star at core-collapse, $v_{\rm esc,cc}$. The wind velocity is then 
\begin{equation}
    v_{\rm w} = v_{\rm esc,cc} \left( \frac{f_{\rm ml}}{0.8}  \right)^{-1}. 
    \label{eq:Vwind}
\end{equation}
Taking $r=v_{\rm w} (t_{\rm cc}-t)$ the expression for the optical depth is  
\begin{equation}
    \tau_{\rm V} =  \intop_{t_{\rm out} } ^{t_{\rm in}} \frac{\kappa_{\rm V} \dot{M}(t)}{4\pi v_{\rm w}^2 (t_{\rm cc} - t)^2}\, {\rm d} t .
    \label{eq:tau}
\end{equation}

\begin{figure*}[ht!]
\centering
\includegraphics[trim= 0cm 0cm 0cm 0cm,clip=true,width=0.97\textwidth]{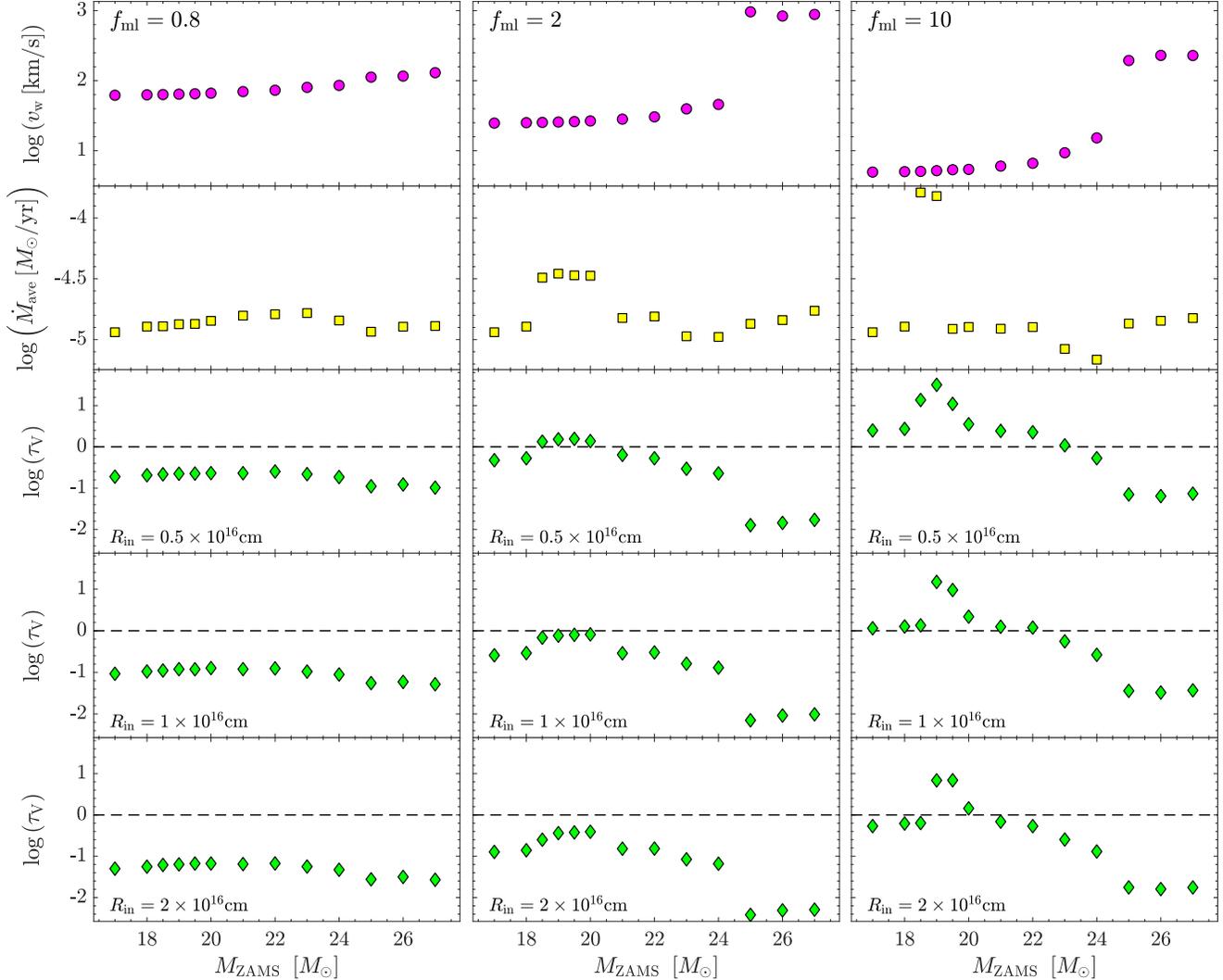} 
\caption{From top row to bottom and in logarithmic scales: The wind velocity according to equation (\ref{eq:Vwind}), the average mass-loss rate in the last $100 \yr$ before core-collapse, and the optical depth of the dust as given by equation (\ref{eq:tau}) with opacity of $\kappa_{\rm V}=100 \cm^2 \g^{-1}$ and for three values of the inner radius, $R_{\rm in}=0.5,~1,~2 \times 10^{16} \cm$; the dashed black line marks: $\tau_{\rm V} = 1$. We calculate each quantity for the 3 instability strip mass-loss scaling factors $f_{\rm ml} = 0.8$ (left column), $f_{\rm ml} =2$ (middle column), and $f_{\rm ml}=10$ (right column). }
\label{fig:Mdot}
\end{figure*}

We present the wind velocity according to equation (\ref{eq:Vwind}) for the different models in the top row of Fig. \ref{fig:Mdot}. In the second row we present the average mass-loss rate in the last 100 years before explosion,  and in the three bottom rows we present the optical depth according to equation (\ref{eq:tau}) for $\kappa_{\rm V} = 100 \cm^2 / \rm{g}$, and for three different values of the inner radius $R_{\rm in} = 0.5,1,2\times10^{16} \cm$. The differences in the optical depth between the three values of the inner radius are very small, and in what follows we will refer to the numbers for $R_{\rm in} = 10^{16} \cm$. We discuss the implications of the optical depth in section \ref{sec:RSGproblem}. 

\begin{figure}
    \centering
    \includegraphics[trim= 0cm 0cm 0cm 0cm,clip=true,width=0.47\textwidth]{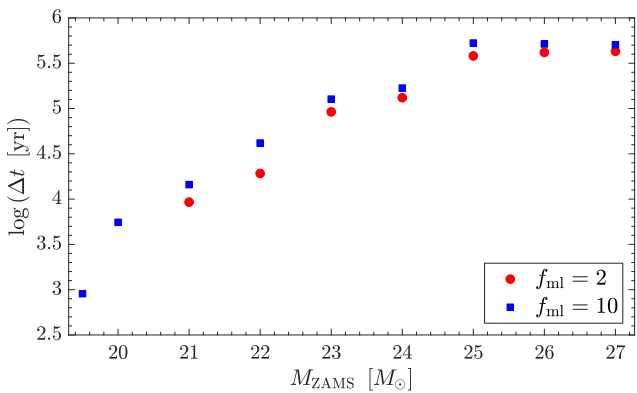}
    \caption{The time of explosion after exiting the instability strip as a function of ZAMS mass.}
    \label{fig:dt}
\end{figure}

Another relevant quantity is the time after the model exists the instability strip and until explosion, which we present in Fig \ref{fig:dt}.
{}

Although \cite{Meynetetal2015} do not consider the RSG problem, it is beneficiary to compare some of our results with theirs. \cite{Meynetetal2015} enhance the mass loss rate by a factor of 10 or 25 when the star is a RSG, which they take to be when the effective temperature is $\log(T_{\rm eff}/K) < 3.7$. Namely, they do not consider an instability strip, but rather they enhance the mass loss rate whenever the star has a low effective temperature. They chose these factors somewhat arbitrarily. Another difference is that we use an initial metalicity of $Z=0.02$ while they take $Z=0.014$.

In evolution without rotation, the $20 M_\odot$ stellar models of \cite{Meynetetal2015} do not have an effective temperature larger than $10^4 \K$ at explosion. We also find that our $20 M_\odot$ do not reach high temperatures at explosion. Their non-rotating model of $25 M_\odot$ reaches a temperature of $\log(T_{\rm eff}/K) \simeq 4.2$ at the end of the evolution, and with a hydrogen mass of $\la 0.1 M_\odot$.  Our $25 M_\odot$ model ends with an effective temperature of $\log(T_{\rm eff}/K) \simeq 4.2$ and with no hydrogen. Generally, for the more massive stars, $M \ga 25 M_\odot$, we obtain different results from those of \cite{Meynetetal2015}. They note that their prescription forms no WC stars (WC are WR stars with strong carbon and oxygen emission lines).
We, on the other hand, account for WR stars, and possibly for WC stars. 
Our results are compatible with the suggestion of \cite{Shenaretal2019} that traditional evolution codes might underestimate mass loss rates during the RSG phase. \cite{Shenaretal2019} based their suggestion on their claim that, at least in the Small Magellanic Cloud, the binary evolution channel does not dominate the formation of WR stars. 
 
Overall, our results and those of \cite{Meynetetal2015} have some similarities, although they are not identical. This is expected because the enhanced mass loss rate prescription is very different. In any case, the results do not contradict each other. The most important similarity is that both studies conclude that with an enhanced mass loss rate a significant fraction of massive stars explode as CCSNe while they are hotter than RSGs. This implies that they might not be SNe II and not even SNe IIb.


\section{Implications to the RSG problem}
\label{sec:RSGproblem}
 
In our search for the mass loss rate that would be necessary to explain the RSG problem for single stars, we increased the mass loss rate in the extension of the instability strip. We described the results of stellar evolution simulations in section \ref{sec:results}. 
 This introduction of high mass-loss rate in the extension of the instability strip splits the stars that enter the strip from right to left to four groups. (1) Stars that explodes while still suffering a very high mass-loss rate and are likely to be IR-bright but visibly faint.  These stars fulfil our request that the enhanced mass loss rate that we introduce forms a CSM that obscures the exploding stars. (2) Stars that leave the strip and explode as SNe II. (3) Stars that leave the strip and explode with hydrogen mass of $0.01 M_\odot \la M_{\rm H, cc} \la 0.5-1M_\odot$ and form SNe IIb. (4) Stars that lose all their hydrogen and explode as SNe Ib. We infer the mass range of each group from Figs. \ref{fig:FinalHandHe} and \ref{fig:Mdot}. 
   
 Because of the large uncertainties in mass-loss rates, boundaries of the extension of the instability strip, and a possible influence by weak binary interaction (our scheme does not treat strong binary interaction), we take the boundaries between the groups as whole solar mass, beside one case. 
  We basically give the boundaries between the above groups for the minimum mass loss rate enhancement that we found to be necessary to account for the RSG problem. 
 \cite{Gordonetal2016} argue that $30 \% - 40\% $ of the yellow supergiants that they study in the galaxies M31 and M33 are likely in a post-RSG phase. The stars that leave the instability strip in our simulations (groups 2-4 above) might account in part for these yellow supergiants.

\subsection{Dust enshrouded IR bright CCSNe}
\label{subsec:IRCCSNe}

From Fig. \ref{fig:Mdot} we see that for $f_{\rm ml}=10$ in our mass-loss scheme this group comprises stars with initial masses of $M_{\rm S,IR} \approx 18.5-20 M_\odot$. We emphasise that the size of the instability strip in these high luminosities is uncertain, and the range might be somewhat larger. 
As well, our scheme refers only to single stars and those that suffer a weak binary interaction. Stars with a strong binary interaction require different calculations. 
In any case, this range of stellar mass ($M_{\rm S,IR} \approx 18.5-20 M_\odot$) is large enough for us to claim that the values of $f_{\rm ml}=10$  (a factor of 12.5 mass loss rate enhancement) is about the minimum value that is required for the mass loss rate enhancement to possibly explain the RSG problem (or part of it).
  
We assume here and below that about half of the stars suffer only weak or no binary interaction. For an initial mass function  (IMF) of $d N \propto M^{-2.35} d M $, we find this group to account for $F_{\rm S,IR} \approx 2\%$ of all CCSNe. With weak binary interactions that enhance mass-loss and somewhat wider instability strip, this group might be $\approx 5 \%$ of all CCSNe.
By a weak binary interaction, we refer to a weak to moderate spin-up by a companion or a weak tidal interaction. Our scheme does not include strong binary interactions where a companion determines the mass-loss rate, e.g., like a massive companion that enters a common envelope.    

In discussing an explosion within a dust shell, we follow \cite{Kochaneketal2012} in treating obscuring by dust. They discuss several important processes, such as the presence of one type of dust, silicate (for massive stars that we study here) or graphitic, and the emission by the dust shell. Since the dust shell is unresolved, its emission adds to the luminosity mainly in the IR. 
The optical depth in the visible of wind with constant velocity $v_{\rm w}$ and a constant mass-loss rate of $\dot M$ is given by equation (\ref{eq:TauV}). In the lower row of Fig. \ref{fig:Mdot} we present the optical depth in the V-band for a dusty wind that takes into account the mass-loss rate variation in our stellar evolution simulations (equation \ref{eq:tau} from an inner radius $R_{\rm in}=10^{16} \cm$), but takes a constant wind velocity (equation \ref{eq:Vwind}). 

Since the shell is not resolved, not all the photons in the visible that are scattered by dust are lost from our beam, and the decrease in the visible light is about a factor of ${\rm few}\times 10$ for $\tau_{\rm V} =5$, or more than three magnitudes in the visible \citep{Kochaneketal2012}.  
 
Shortly after the explosion the SN ejecta collides with the dense wind, the CSM. The interaction of the ejecta with the CSM converts kinetic energy to radiation. We scale the efficiency of this process to be $\epsilon_i=0.1$ and the shock velocity into the CSM to be $v_{\rm s}=4000 \km \s^{-1}$ (e.g.,  \citealt{Foxetal2013, Foxetal2015}) 
\begin{equation}
\begin{aligned}
& L_i= \epsilon_i \dot M \frac{v^3_{\rm s}} {2 v_{\rm w}}
= 5.3 \times 10^6   
\left( \frac {\dot M}{10^{-4} M_\odot \yr^{-1}} \right)
 \\ & \times
\left( \frac {v_{\rm s}}{4000 \km \s^{-1}} \right)^{3}
\left( \frac {v_{\rm w}}{10 \km \s^{-1}} \right)^{-1} 
\left( \frac {\epsilon_i}{0.1} \right) L_\odot . 
\end{aligned}
\label{eq:Li}
\end{equation}
This corresponds to a bolometric magnitude of about $-12$, fainter by several magnitudes relative to typical CCSNe. In addition, the dust that still resides at large distances will make the SN redder, and so the visual magnitude will be lower even relative to typical CCSNe.  
Such events might be classified at first place as intermediate luminosity optical transients (ILOTs), rather than CCSNe. But they are fainter in the visible and therefore will be detected in much lower numbers than CCSNe that are not enshrouded by a dense dusty wind.  

We conclude that the dusty wind reduces the luminosity in the visible by several magnitudes. Present observations can still detect such type II CCSNe, but at much smaller numbers than their occurrence rate.  
As we write above, these are only for stars in the initial mass range of $M_{\rm S,IR} \approx 18.5-20 M_\odot$. 

\subsection{Type II CCSNe}
\label{subsec:TypeII}
 
This group is of stars that have hydrogen mass at core-collapse of $M_{\rm H, cc} \gtrsim 1 M_\odot$, and that are not enshrouded by optically thick dust. From Fig.  \ref{fig:FinalHandHe} we find the upper mass of this group and from Fig. \ref{fig:Mdot} its lower mass. These give for the mass range of this group $M_{\rm S,II} \simeq 20-21 M_\odot$. This mass range amounts to $\approx 2\%$ of all CCSNe, or $F_{\rm S,II} \approx 1 \%$ of all CCSNe if we take those that do not suffer strong binary interaction.  

\subsection{Type IIb CCSNe}
\label{subsec:TypeIIb}

SNe~IIb are CCSNe that in the first several days have strong hydrogen lines, but later these lines substantially weaken and even disappear. This results from low hydrogen mass at explosion, about $M_{\rm H, cc} \simeq 0.03-0.5 M_\odot$ (e.g., \citealt{Meynetetal2015, Yoonetal2017}), or even up to $M_{\rm H,env} \le 1 M_\odot$ (e.g., \citealt{Sravanetal2018}). SNe~IIb make $f_{\rm IIb,H} \simeq 10-12 \%$ of all CCSNe in high metallicity stellar populations (e.g., \citealt{Sravanetal2018}).
From Fig. \ref{fig:FinalHandHe} we find that the relevant mass range for SNe IIb progenitors in our $f_{\rm ml} =10$ case is $M_{\rm S,IIb} \simeq 21-24 M_\odot$. \cite{Meynetetal2015} also find that with their scheme of RSG enhanced mass loss rate many stars end with low hydrogen mass. For an IMF of $d N \propto M^{-2.35} dM$ this amounts to $\simeq 0.045$ of all CCSNe. However, if about half of these stars suffer strong binary interaction that our scheme does not consider, the single-star and weak binary interaction channels that we study here for SNe~IIb correspond to $F_{\rm S,IIb} \approx 2\%$ of all CCSNe. We note that \cite{Naimanetal2020} crudely suggest that the single-star channel for SNe IIb accounts for $\approx 2-4\%$ of all CCSNe (about $20-40 \%$ of all SNe~IIb).  
  
\subsection{Type Ib CCSNe}
\label{subsec:TypeIb}
 
In the mass range we calculate here this group comes from stars with an initial mass of $M_{\rm S,Ib} \gtrsim 24-25$, as we see from Fig. \ref{fig:FinalHandHe}.  \cite{Meynetetal2015} simulated enhanced mass loss rate from stars up to $M_{\rm ZAMS}=25 M_\odot$ and find that all of them maintain hydrogen. We find this limit at $M_{\rm ZAMS}=24 M_\odot$. Consider that we do not use the same mass loss rate enhancement scheme as they do, this is not a large difference. This range amounts to $\simeq 20 \%$ of all CCSNe if we take the upper mass limit to be $M_{\rm ZAMS}=100 M_\odot$. If we consider that about half suffer strong binary interaction  (e.g., \citealt{Sanaetal2012}), the single star evolution (including weak binary interactions) that we study here amounts to $F_{\rm S,Ib} \approx 10 \%$ of all CCSNe. Some of them might lose also all their helium and lead to SNe~Ic.      
Our finding that most, $\simeq 2/3$, of the stars with $M_{\rm S,Ib} \gtrsim 18$ form SNe~Ib and possibly SNe~Ic, is compatible with the finding of \cite{Smartt2015R}. The claim of \cite{Stritzingeretal2020} that the progenitor of the SN Ib  LSQ13abf had an initial mass of $\ga 25 M_\odot$ (or $ \ga 20-25 M_\odot$ in an alternative model) supports our claim. 

This group of stars adds to the role of the mass loss rate enhancement in accounting for the RSG problem. Namely, the research question of this study, which is about the enhanced mass loss rate that is necessary to explain the RSG problem, refers both to obscured CCSNe and to transforming some RSG stars to progenitors of SNe Ibc (section \ref{sec:intro}).

The finding above has implications to the rate of formation of WR stars in the single-star channel. Many WR stars are observed to be single, but traditional stellar evolution calculations are short in accounting for these WR stars as well as other properties (e.g., \citealt{Shenaretal2020}). For that, some (e.g., \citealt{SchootemeijerLanger2018S}) claim that the companion in many of these systems is a low mass star that observations did not reveal yet. However, other studies (e.g., \citealt{Shenaretal2016}) suggest that, at least in the Small Magellanic Cloud, the binary evolution channel does not dominate WR formation. \cite{Shenaretal2019} study WR stars and their formation in the binary and single-stellar channels, and suggest that it is possible that traditional evolution codes underestimate mass loss (mainly) during the red supergiant phase (for an earlier similar claim see \citealt{Vanbevereneta1998a, Vanbevereneta1998b}.
Our assumption of an enhanced mass loss rate in the upper instability strip, and the results of high fraction of SNe Ibc progenitors, are compatible with the claim of \cite{Shenaretal2019} of a higher mass loss rate than what traditional evolution code give. 
 
\section{Summary}
\label{sec:summary}

We are motivated by the theoretical disagreement on the fate of star with ZAMS mass of $M_{\rm ZAMS} \gtrsim 18M\odot$ (section \ref{sec:intro}). 
 For that, we examined the question of what enhanced mass loss rate during the RSG would be necessary to explain the RSG problem. We noted that there is no support from observations for such a large enhanced mass loss rate factor. Therefore, we raised a theoretical question. We found that we need to increase the mass loss rate by at least a factor of about ten (we used a factor of 12.5 in this study) in the instability strip of the RSG to explain the RSG problem (or part of it). 
   
Using the numerical stellar evolution code \textsc{mesa} we have simulated the evolution of 48 stellar models to the point of core-collapse and explored the effect of an enhanced mass-loss inside the instability strip as the evolved stars cross from right to left at very high luminosities. Based on \cite{YoonCantiello2010} we assumed an enhanced mass-loss rate as the star crosses the instability strip from right to left at high luminosities (grey area of the instability strip on Fig. \ref{fig:HRdif}). 
Our mass-loss prescription is for single star evolution and possibly weak binary interaction.  We do not include strong binary interaction. 

We concentrated on two pre-core-collapse stellar properties, the stellar hydrogen mass (Fig. \ref{fig:FinalHandHe}), and the optical depth of the dusty wind (Fig. \ref{fig:Mdot}). From these properties we divide the stars that enter or cross the upper (extension) instability strip to four groups with very uncertain mass boundaries between them.  (1) Stars that explode as SNe II while they are in the strip and therefore are enshrouded by dust (section \ref{subsec:IRCCSNe}). These have initial mass in the range of  $M_{\rm S,IR} \simeq 18.5-20 M_\odot$. (2) Stars that leave the instability strip from the left and explode as SNe II. They have  $M_{\rm S,II} \simeq 20-21 M_\odot$ (section \ref{subsec:TypeII}). (3) Stars that leave the strip and at core-collapse have a hydrogen mass of $M_{\rm H, cc} \la 0.5-1 M_\odot$. They explode as SNe IIb and have $M_{\rm S,IIb} \simeq 21-24 M_\odot$ (section \ref{subsec:TypeIIb}). (4) Stars that leave the strip and explode as SNe Ib and possibly as SNe Ic. These have $M_{\rm S,Ib} \gtrsim 24$ (section \ref{subsec:TypeIb}). 

In short, we have found that an enhanced mass loss rate in the instability strip (Fig. \ref{fig:HRdif}) by a factor of about ten or more, helps solving the RSG problem by causing some CCSN to be obscured by dust (group 1 above) and by causing other stars to explode as SNe Ibc (group 4 above).
       
Because the mass boundaries of the four groups are highly uncertain, so are the fraction $F_{\rm S}$ of each group is highly uncertain.
For the minimum mass loss rate enhancement that would be necessary to solve the RSG problem, our estimated fractions of CCSNe in each of these four groups are $F_{\rm S,IR} \approx 2 \%$, 
$F_{\rm S,II} \approx 1 \%$,  $F_{\rm S,IIb} \approx 2\%$, and $F_{\rm S,Ib} \approx 10\%$, respectively.
In estimating these fractions we used the IMF of $dN \propto M^{-2.35} dM$ with a maximum mass of $100 M_\odot$ and assumed that about half of the stars suffer strong binary interaction that we do not consider here. 
Therefore, our assumption of enhanced mass-loss while in the instability strip implies that single star evolution brings only a fraction of  
\begin{equation}
\eta_{\rm S,II} \equiv \frac{ F_{\rm S,II} +  F_{\rm S,IIb}} { F_{\rm S,IR} +  F_{\rm S,II} +  F_{\rm S,IIb} + F_{\rm S,Ib}} \approx 20 \% 
\label{eq:eta}
\end{equation}
of stars with $M_{\rm ZAMS} \gtrsim 18.5 M_\odot$ to end as SNe~II or SNe~IIb that are not heavily enshrouded by dusty CSM.  

\cite{Smartt2015R} lists 30
progenitors of SN type II or IIb which all have a ZAMS mass of $M_{\rm ZAMS} \lesssim 18M_\odot$. From that he argues that the IMF implies that if all theses stars explode there should be $\approx 13$ CCSNe of types II and IIb with a progenitor of $M_{\rm ZAMS} \gtrsim 18M_\odot$. According to our analysis (equation \ref{eq:eta}) we expect that out of these $13$ SNe with progenitor mass $M_{\rm ZAMS} \gtrsim 18M_\odot$, only $\approx 2-3$ are SNe II or IIb (and binary interaction can reduce this number further by forming more SNe Ibc)\footnote{At the Symposium "The Deaths and Afterlives of Stars" (Space Telescope Science Institute, April 22-24, 2019) Smartt updated the observed number of progenitors with $M_{\rm ZAMS} \lesssim 18M_\odot$ to 35 ( \url{https://cloudproject.hosted.panopto.com/Panopto/Pages/Viewer.aspx?id=d6217958-cb9c-4825-8005-aa3700f5dfb0}). In this case 15 progenitors with a mass of $M_{\rm ZAMS} \gtrsim 18 M_\odot$ are expected. By our analysis, only 3 of them should be type II or IIb CCSNe.
}. 

Our main conclusion is that the statistical uncertainties are too large to decide whether many stars with $M_{\rm ZAMS} \gtrsim 18M_\odot$ do not explode as expected in the neutrino driven explosion mechanism, or whether most of them form SNe Ibc and obscured SNe II that are IR-bright, as expected by the jittering jets explosion mechanism.

\section*{Acknowledgements}

We thank an anonymous referee for detailed and useful comments. This research was supported by a grant from the Israel Science Foundation.
N.S. research is partially supported by the Charles Wolfson Academic Chair.

\label{lastpage}
\end{document}